\documentclass[sigconf, nonacm]{acmart}

\usepackage{booktabs}
\usepackage{graphicx}
\usepackage{subcaption}
\newcommand\vldbdoi{XX.XX/XXX.XX}

\newcommand\vldbavailabilityurl{https://github.com/UIC-InDeXLab/RSR-core}
\begin{document}
\title{RSR-core: A High-Performance Engine for Low-Bit Matrix-Vector Multiplication}

%%
%% The "author" command and its associated commands are used to define the authors and their affiliations.
\author{Mohsen Dehghankar}
\affiliation{%
  \institution{University of Illinois Chicago}
  % \streetaddress{P.O. Box 1212}
  % \state{Ireland}
}
\email{mdehgh2@uic.edu}

\author{Abolfazl Asudeh}
% \orcid{0000-0002-1825-0097}
\affiliation{%
  \institution{University of Illinois Chicago}
  % \streetaddress{1 Th{\o}rv{\"a}ld Circle}
  % \city{Hekla}
  % \country{Iceland}
}
\email{asudeh@uic.edu}

\begin{abstract}
Matrix–vector multiplication is a fundamental building block in neural networks, vector databases, and large language models, particularly during inference. As a result, efficient matrix–vector multiplication engines directly translate into more efficient inference.

Recent work has explored low-bit quantization of model weights, where matrices are represented using binary (1-bit) or ternary (1.58-bit) values while activation kept in higher precision. These representations enable efficient hardware-level computation. In parallel, algorithms such as Redundant Segment Reduction (RSR) provide theoretical guarantees for accelerating low-bit matrix–vector multiplication. However, existing implementations operate at the application level and cannot be efficiently integrated into hardware kernels, limiting practical performance.

To bridge this gap, we present RSR-core, a high-performance engine that implements the RSR algorithm as optimized low-level kernels for both CPU and CUDA environments. RSR-core supports efficient matrix–vector multiplication for binary and ternary weight matrices and general vectors while enabling practical deployment of RSR algorithm in real inference pipelines.

RSR-core is provided as a production-ready engine with HuggingFace integration for preprocessing low-bit models and running accelerated inference.

Experimental results demonstrate significant performance improvements over baseline HuggingFace PyTorch multiplication, achieving up to 62× speedup on CPU and up to 1.9× speedup for token generation on CUDA for popular ternary LLMs. The source code is publicly available at \href{https://github.com/UIC-InDeXLab/RSR-core}{\color{blue}RSR-core repository\color{black}}\footnote{\href{https://github.com/UIC-InDeXLab/RSR-core}{https://github.com/UIC-InDeXLab/RSR-core}}.

\end{abstract}

\maketitle

% \pagestyle{\vldbpagestyle}
% \begingroup\small\noindent\raggedright\textbf{PVLDB Reference Format:}\\
% \vldbauthors. \vldbtitle. PVLDB, \vldbvolume(\vldbissue): \vldbpages, \vldbyear.\\
% \href{https://doi.org/\vldbdoi}{doi:\vldbdoi}
% \endgroup
% \begingroup
% \renewcommand\thefootnote{}\footnote{\noindent
% This work is licensed under the Creative Commons BY-NC-ND 4.0 International License. Visit \url{https://creativecommons.org/licenses/by-nc-nd/4.0/} to view a copy of this license. For any use beyond those covered by this license, obtain permission by emailing \href{mailto:info@vldb.org}{info@vldb.org}. Copyright is held by the owner/author(s). Publication rights licensed to the VLDB Endowment. \\
% \raggedright Proceedings of the VLDB Endowment, Vol. \vldbvolume, No. \vldbissue\ %
% ISSN 2150-8097. \\
% \href{https://doi.org/\vldbdoi}{doi:\vldbdoi} \\
% }\addtocounter{footnote}{-1}\endgroup
% %%% VLDB block end %%%

% %%% do not modify the following VLDB block %%
% %%% VLDB block start %%%
% \ifdefempty{\vldbavailabilityurl}{}{
% \vspace{.3cm}
% \begingroup\small\noindent\raggedright\textbf{PVLDB Artifact Availability:}\\
% The source code, data, and/or other artifacts have been made available at \url{\vldbavailabilityurl}.
% \endgroup

\section{Introduction}

\begin{figure}[t]
  \centering
  \includegraphics[width=\linewidth]{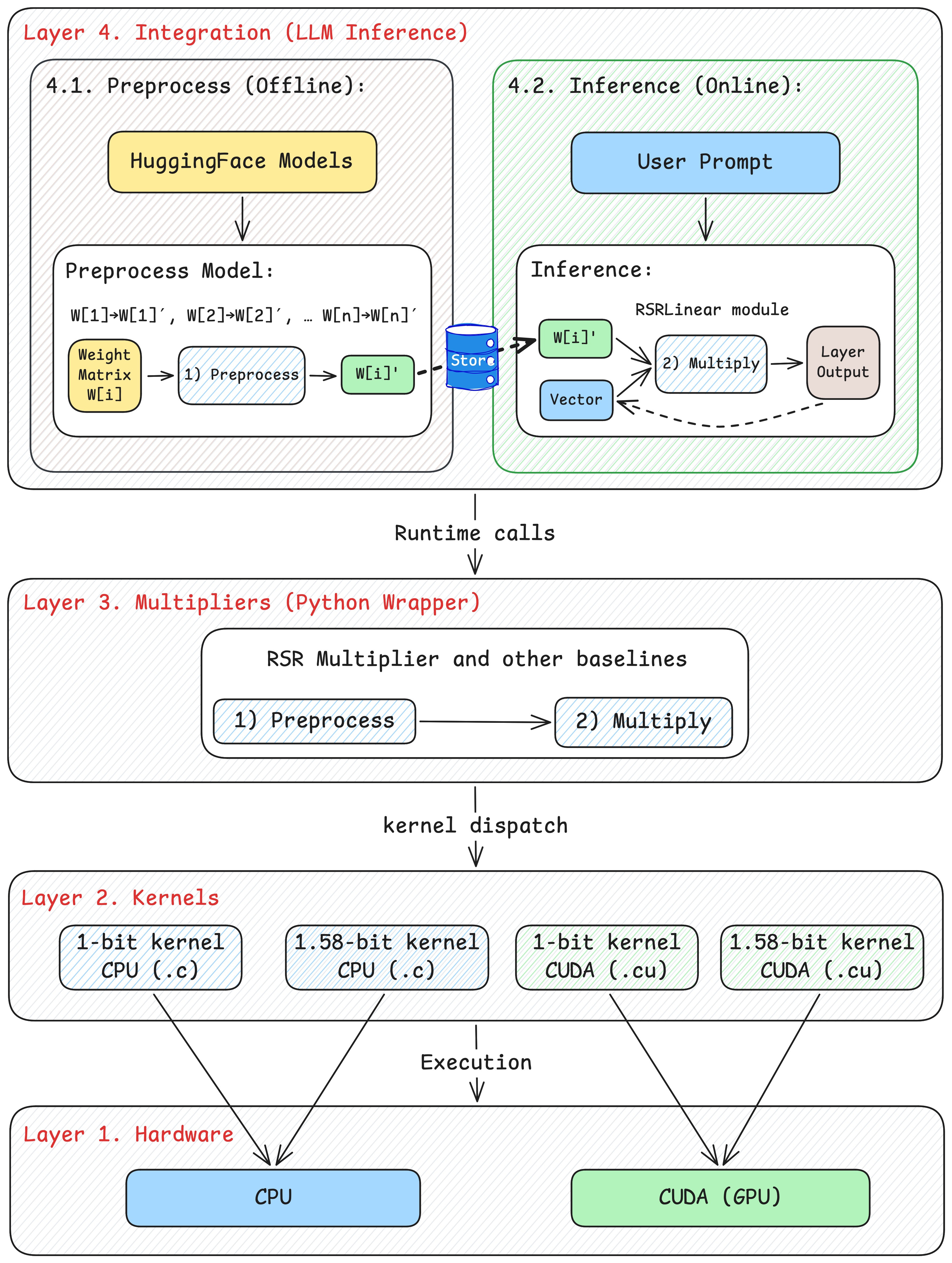}
  \caption{The layered architecture of RSR-core system.}
  \label{fig:architecture}
\end{figure}

% problem and context
Inference is a critical stage in neural networks, particularly for large language models (LLMs), where it dominates both energy consumption and resource allocation~\cite{zhou2024survey, miao2025towards}. A key computational bottleneck during inference is matrix--vector multiplication $Mv$, where the matrix $M$ represents fixed model weights after training and the vector $v$ corresponds to input activations generated during execution. Improving the efficiency of matrix--vector multiplication therefore directly translates into faster and more resource-efficient neural model inference.

% relation to vldb
Efficient matrix--vector multiplication is also central to many data management and retrieval workloads beyond neural inference~\cite{douze2025machine, chen2025maximum}. For example, vector database systems compute inner products between a query vector and indexed data points for nearest neighbor retrieval, where the indexed dataset can be viewed as a matrix and the query as the input vector~\cite{douze2024faiss}. Similarly, ranking over tabular datasets applies user-defined weight vectors to attribute columns to compute scores for sorting and filtering, where the table rows form the matrix and the weight vector defines the ranking function~\cite{rolleke2006general, trabelsi2021neural}.

% Existing approaches and gap
One common approach to accelerating matrix--vector multiplication, specifically in model inference, is low-bit quantization of weight matrices~\cite{ma2024era, wang2023bitnet}. In low-bit quantized LLMs, matrices are represented using binary or ternary values while activation vectors remain in higher precision, enabling specialized hardware-efficient implementations that reduce computation cost during inference.

Orthogonal to hardware-aware quantization approaches, algorithmic methods aim to further accelerate matrix--vector multiplication. In particular, \emph{Redundant Segment Reduction (RSR)}~\cite{dehghankar2024efficient} reduces the computational cost of low-bit matrix--vector multiplication by a logarithmic factor by exploiting redundancy in matrix structure. Specifically, identical column patterns\footnote{In the original paper, the multiplication $v \cdot M$ is considered instead of $M v$, and therefore rows (rather than columns) are grouped together.} contribute identical partial results and can therefore be aggregated before multiplication, reducing the total number of required operations \emph{without affecting inference accuracy}. However, existing implementations of RSR operate at the application level and cannot be efficiently integrated into hardware kernel-level execution, limiting their practical performance benefits in real inference systems.

% Our work
In this work, we address this gap by providing a kernel-level implementation of the RSR algorithm and enabling its deployment in practical inference pipelines. To the best of our knowledge, this is the first implementation of RSR optimized for both CPU and CUDA inference environments.

% Our system and contribution
We present \textbf{RSR-core}, a high-performance engine implementing RSR as optimized CPU and CUDA kernels for efficient low-bit matrix--vector multiplication. RSR-core provides a production-ready Python interface integrated with HuggingFace workflows~\cite{huggingface}, enabling users to preprocess quantized models once and execute accelerated inference efficiently through an end-to-end workflow.

We benchmark RSR-core against widely used hardware-optimized matrix multiplication implementations provided by the PyTorch~\cite{paszke2019pytorch} backend and used within the HuggingFace inference interface, as well as BitNet~\cite{wang2023bitnet}. Experimental results show substantial performance improvements, achieving up to $62\times$ speedup on CPU and up to $1.9\times$ speedup on CUDA for popular ternary LLM inference workloads compared to PyTorch \texttt{bfloat16} computation, while preserving exact inference accuracy.

\section{System Overview}

% Problem definition
We consider matrix--vector multiplication $M \cdot v$, where $M$ is a binary matrix with entries in $\{0, 1\}$ or a ternary matrix with entries in $\{-1, 0, +1\}$. The matrix corresponds to quantized model weights and remains fixed after preprocessing, while the vector $v$ represents input activations arriving during inference, as discussed in the Introduction section. The goal is to accelerate this multiplication in practical inference pipelines while preserving exact computation.

\vspace{1mm}
\noindent{\bf Algorithm overview.}
The Redundant Segment Reduction (RSR) algorithm~\cite{dehghankar2024efficient} consists of two phases: an offline preprocessing applied once to the matrix $M$, followed by a fast online multiplication phase applied repeatedly during inference. During \emph{preprocessing}, RSR splits the matrix into horizontal blocks of $k$ rows and identifies identical column segments within each block, grouping columns that share the same bit pattern. The preprocessing stage produces auxiliary data structures---including column permutations, group boundaries, and scatter indices---that enable efficient reuse of intermediate results during inference.
During \emph{inference}, given an input vector $v$, RSR aggregates vector entries corresponding to identical column segments before multiplication and distributes contributions only once per unique segment. This reduces the number of arithmetic operations required for computing $M \cdot v$ while preserving exact multiplication results without sacrificing the space usage.

\subsection{System Architecture}

% System architecture
RSR-core is an optimized RSR execution engine using a layered architecture illustrated in Figure~\ref{fig:architecture}. At its core, the system provides kernel-level implementations of RSR multiplication on both CPU (C kernels) and CUDA (GPU kernels). These kernels are exposed through intermediate multiplier abstractions and integrated into a Python interface that supports end-to-end preprocessing and inference workflows.

The architecture enables users to preprocess quantized models once and reuse the generated artifacts across repeated inference calls. The full source code is publicly available at
\href{https://github.com/UIC-InDeXLab/RSR-core}{\color{blue} this repository\color{black}}.

\subsection{Kernels Implementation}

While the RSR algorithm is conceptually straightforward, a direct Python implementation does not yield practical speedups due to interpreter overhead, general-purpose sorting, and inefficient memory access patterns during the gather, aggregate, and scatter phases. RSR-core therefore provides native kernel implementations in C (CPU) and CUDA (GPU) for both binary (1-bit) and ternary (1.58-bit) matrices. A key design choice spans all backends: preprocessing uses \emph{counting sort} over the discrete pattern space ($2^k$ buckets for binary, $4^k$ for ternary), achieving $O(n + \text{buckets})$ complexity per block instead of $O(n \log n)$ from comparison-based sorting. Metadata is uniformly shrunk to compact types---permutation indices and group boundaries stored as 16-bit integers, and scatter targets encoded as fixed-size bitmasks rather than variable-length index arrays---reducing bandwidth pressure in the inference hot path.

On the CPU, the C kernels fuse the gather and aggregate phases into a single pass over the input vector, accumulating partial sums directly without materializing intermediate arrays. The binary kernel uses scalar-unrolled loads with software prefetch hints, which outperforms hardware vector gather instructions on the random-access pattern inherent to RSR. For ternary matrices, where each group must scatter both a positive and a negative contribution, the kernel replaces variable-length signed scatter lists with two compact bitmasks per group and iterates their set bits directly, making the per-group scatter cost independent of how many output rows are active. All CPU kernels are parallelized across row blocks.

For end-to-end model inference on CPU, RSR-core further fuses activation quantization with the RSR matrix-vector multiply into a single native call, and batches sibling linear layers that share the same input vector (e.g., $W_q$, $W_k$, $W_v$) into one combined operation. This eliminates repeated Python dispatch and redundant quantization, and exposes a larger pool of row blocks to distribute across cores---a layer of optimization that contributes substantially to the observed wall-clock speedups beyond the kernel itself.

On CUDA, each row block is assigned to one thread block, and warps within a block process groups in parallel. Each group's metadata is packed into a single 64-bit word encoding the permutation range, length, and scatter masks, so the kernel loads one value per group and begins work immediately. Permutation indices are sorted within each group during preprocessing, which does not affect the computed sum but improves cache locality on the gather reads. Per-warp shared-memory partial buffers avoid contention on output writes, and zero-contribution groups are dropped during preprocessing to eliminate useless work. Compile-time specializations on $k$ allow the compiler to fully unroll scatter loops for each supported block height\footnote{See the code repository for more technical implementation detail.}.

\subsection{Multipliers}

% wrappers
On top of the kernel implementations, RSR-core provides a set of Python classes called \texttt{Multiplier}. These classes encapsulate multiplier-specific logic, including both RSR-based multiplication and baseline multipliers such as BitNet. Each multiplier exposes two primary operations: preprocessing and multiplication.

The preprocessing stage operates on a given low-bit matrix and produces reusable artifacts that encode segment-level structure for efficient inference-time execution. The multiplication stage then uses these artifacts together with an input vector $v$ to compute the exact matrix--vector product $M \cdot v$.\footnote{Note that preprocessing is required only for RSR-based multipliers.}

\subsection{Integration}

% Inference integration and user interaction
RSR-core provides an interface for executing inference with ternary large language models and integrates directly with models available through the HuggingFace model hub\footnote{\href{https://huggingface.co/models}{https://huggingface.co/models}}. Models that use ternary (BitLinear) layers, such as BitNet~\cite{wang2023bitnet} and its variants, can be used with RSR-core without modifying existing inference pipelines\footnote{For example, \texttt{microsoft/bitnet-b1.58-xxx} models.}.

Users apply preprocessing once to a selected model downloaded from the HuggingFace model hub. The generated preprocessing artifacts are stored locally, with at most the same space usage as the models themselves, and reused for subsequent inference executions. During inference, users can interact with the model through standard prompting workflows and obtain responses using accelerated matrix--vector multiplication provided by RSR-core (see demonstration section).

RSR-core also includes a lightweight user interface that simplifies model selection, preprocessing management, storage monitoring of generated artifacts, and interactive prompting of the models.

Integration with PyTorch-based inference pipelines is provided through a custom \texttt{PyTorch} module called \texttt{RSRLinear}, which replaces ternary linear layers with RSR-enabled implementations. \emph{Any system using PyTorch} for neural network inference can incorporate RSR-core through this module with minimal changes.

\begin{figure}[t]
    \centering
    \includegraphics[width=0.49\linewidth]{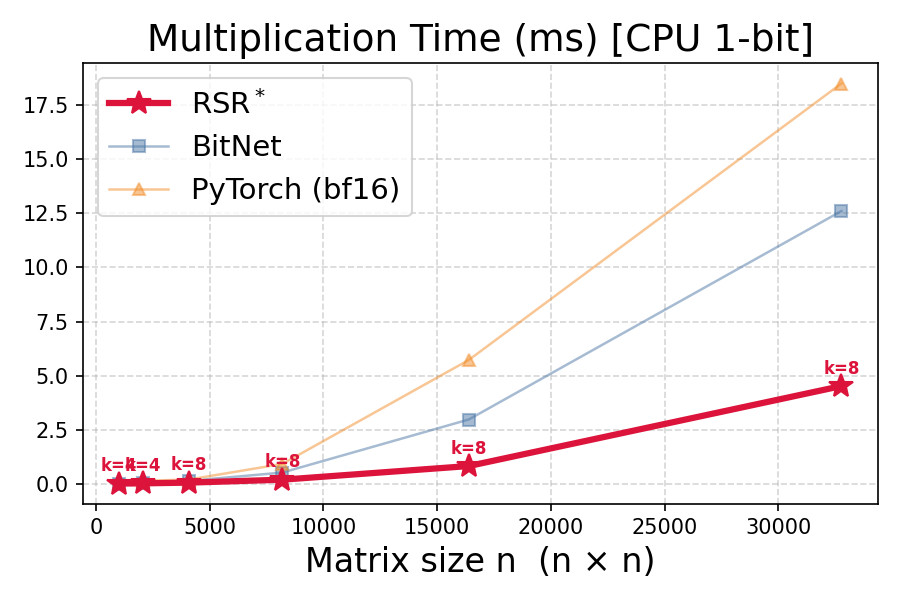}
    \includegraphics[width=0.49\linewidth]{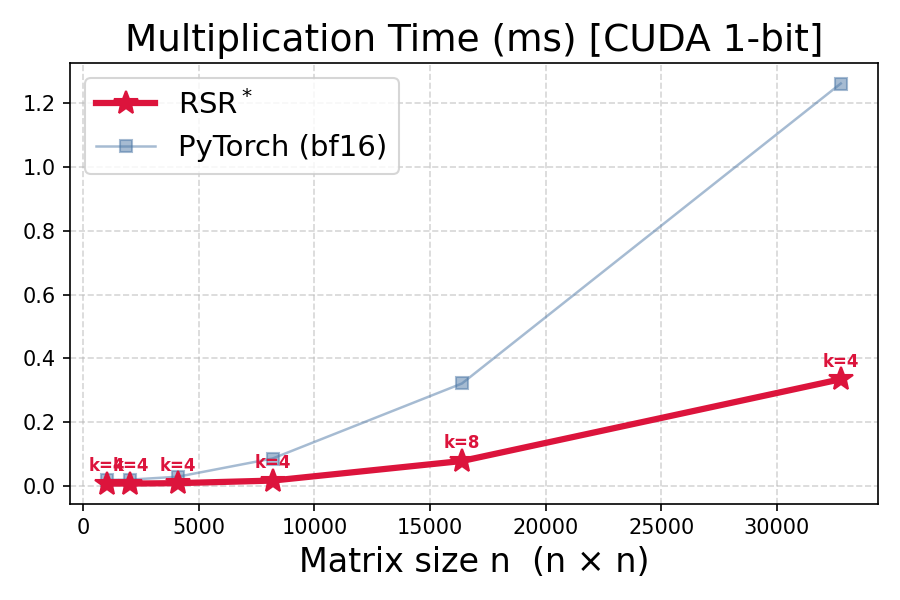}
    \caption{Benchmarking the matrix multiplication on CUDA and CPU. Matrix is binary while vector is \texttt{float32}.}
    \label{fig:mm_benchmark}
\end{figure}

\begin{table}[t]
\centering
\small
\caption{Ternary (1.58-bit) LLM Inference Speedup over HuggingFace PyTorch \texttt{bfloat16}}
\label{tab:token_benchmark}
\begin{tabular}{@{}l r r r@{}}
\toprule
\textbf{Model} & \textbf{HF (Tok/s)} & \textbf{RSR (Tok/s)} & \textbf{Speed-up} \\
               % & \textbf{Tok/s} & \textbf{Tok/s} & \textbf{up} \\
\midrule
\multicolumn{4}{l}{\textit{CPU}} \\
\midrule
Falcon3-10B-1.58b   & 0.2  & \textbf{11.3} & \textbf{62.0x} \\
Llama3-8B-1.58b     & 0.2  & \textbf{13.4} & \textbf{53.8x} \\
BitNet-2B-4T-bf16   & 2.1  & \textbf{28.8} & \textbf{13.9x} \\
BitNet-2B-4T        & 14.2 & \textbf{29.3} & \textbf{2.1x}  \\
\midrule
\multicolumn{4}{l}{\textit{CUDA}} \\
\midrule
Falcon3-10B-1.58b   & 25.2 & \textbf{47.4} & \textbf{1.9x} \\
Llama3-8B-1.58b     & 31.9 & \textbf{59.3} & \textbf{1.9x} \\
BitNet-2B-4T-bf16   & 33.1 & \textbf{57.4} & \textbf{1.7x} \\
BitNet-2B-4T        & 41.6 & \textbf{57.1} & \textbf{1.4x} \\
\bottomrule
\end{tabular}
\end{table}

\subsection{Benchmarking Results}

% benchmark results
To evaluate the performance of RSR-core, we compare multiplication time against baseline implementations provided by PyTorch, which already use hardware-optimized kernels on both CPU and CUDA platforms. We also compare against BitNet~\cite{wang2023bitnet} as an additional baseline for low-bit matrix multiplication. The results (Figure~\ref{fig:mm_benchmark}) show substantial speedups for matrix--vector multiplication on both CPU and CUDA.

We further evaluate end-to-end LLM inference performance during the decoding (token generation) phase and compare against HuggingFace PyTorch inference on both CPU and CUDA platforms. The results (Table~\ref{tab:token_benchmark}) demonstrate significant improvements in token generation throughput using RSR-core. Additional benchmarking results are available in the project repository.

\begin{figure*}[t]
  \centering
  \begin{subfigure}[t]{0.48\textwidth}
    \includegraphics[width=\textwidth]{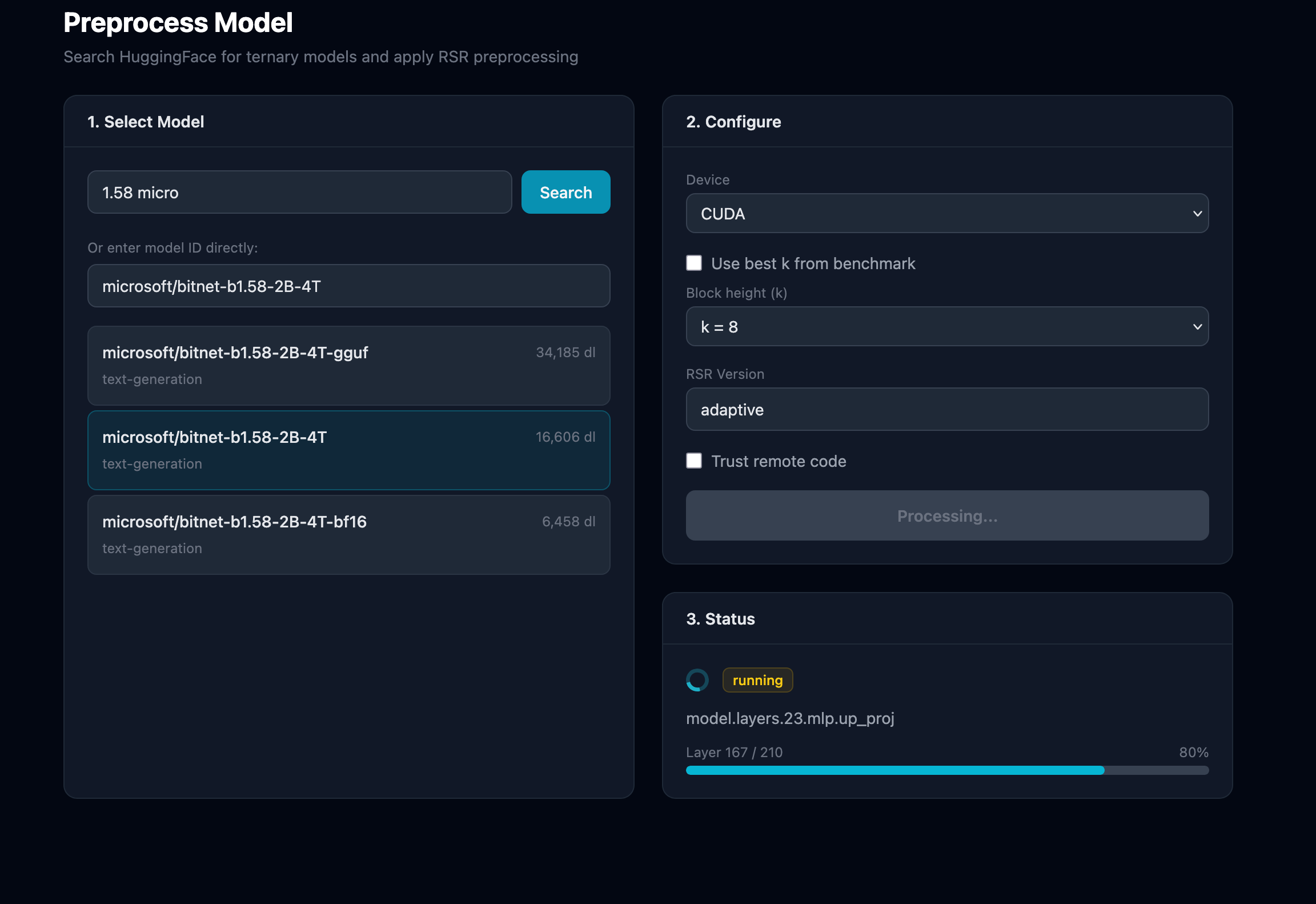}
    \caption{Model preprocessing.}
    \label{fig:ui_preprocessing}
  \end{subfigure}
  \hfill
  \begin{subfigure}[t]{0.48\textwidth}
    \includegraphics[width=\textwidth]{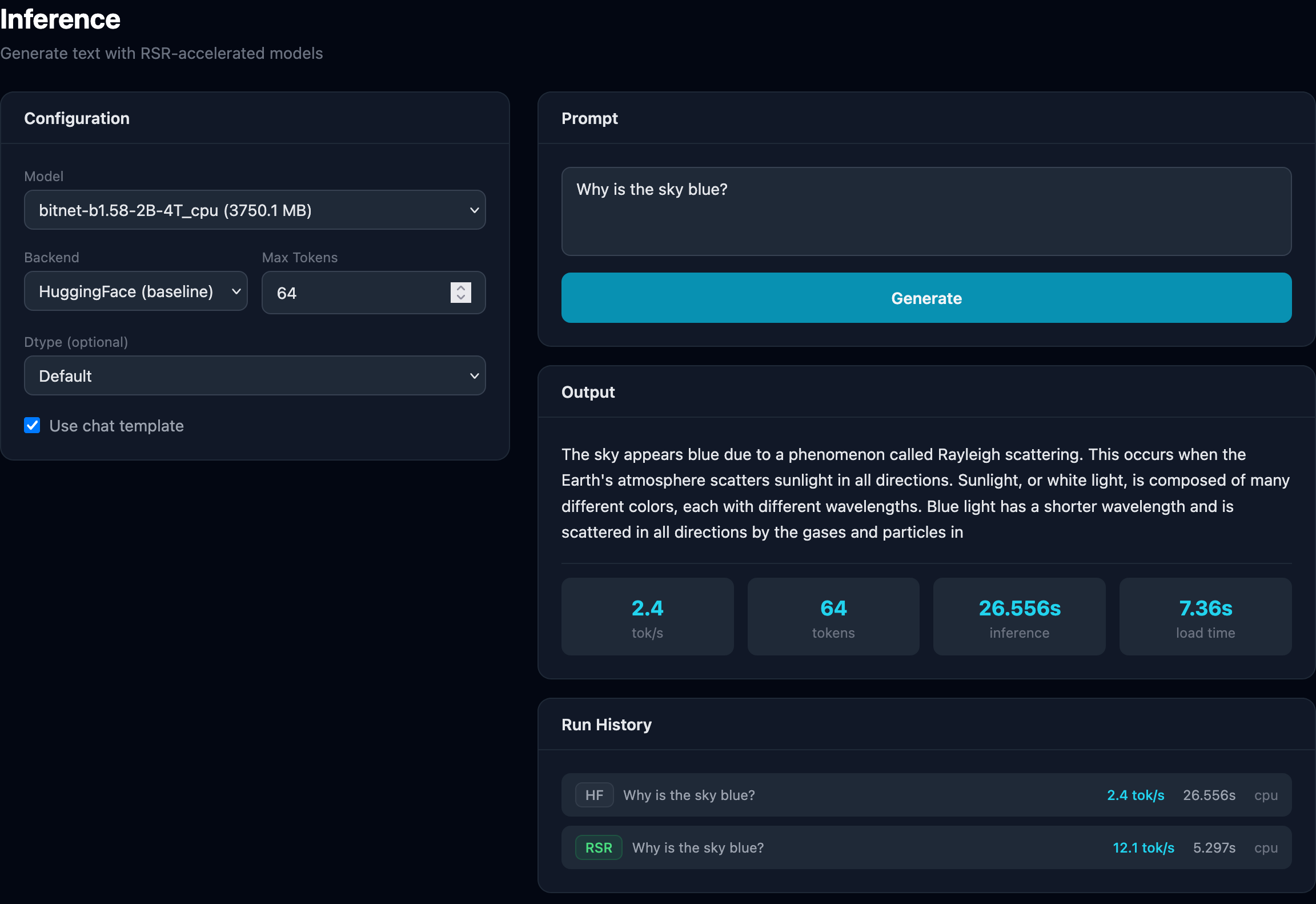}
    \caption{Model inference.}
    \label{fig:ui_inference}
  \end{subfigure}

  \vspace{0.5em}

  \begin{subfigure}[t]{0.48\textwidth}
    \includegraphics[width=\textwidth]{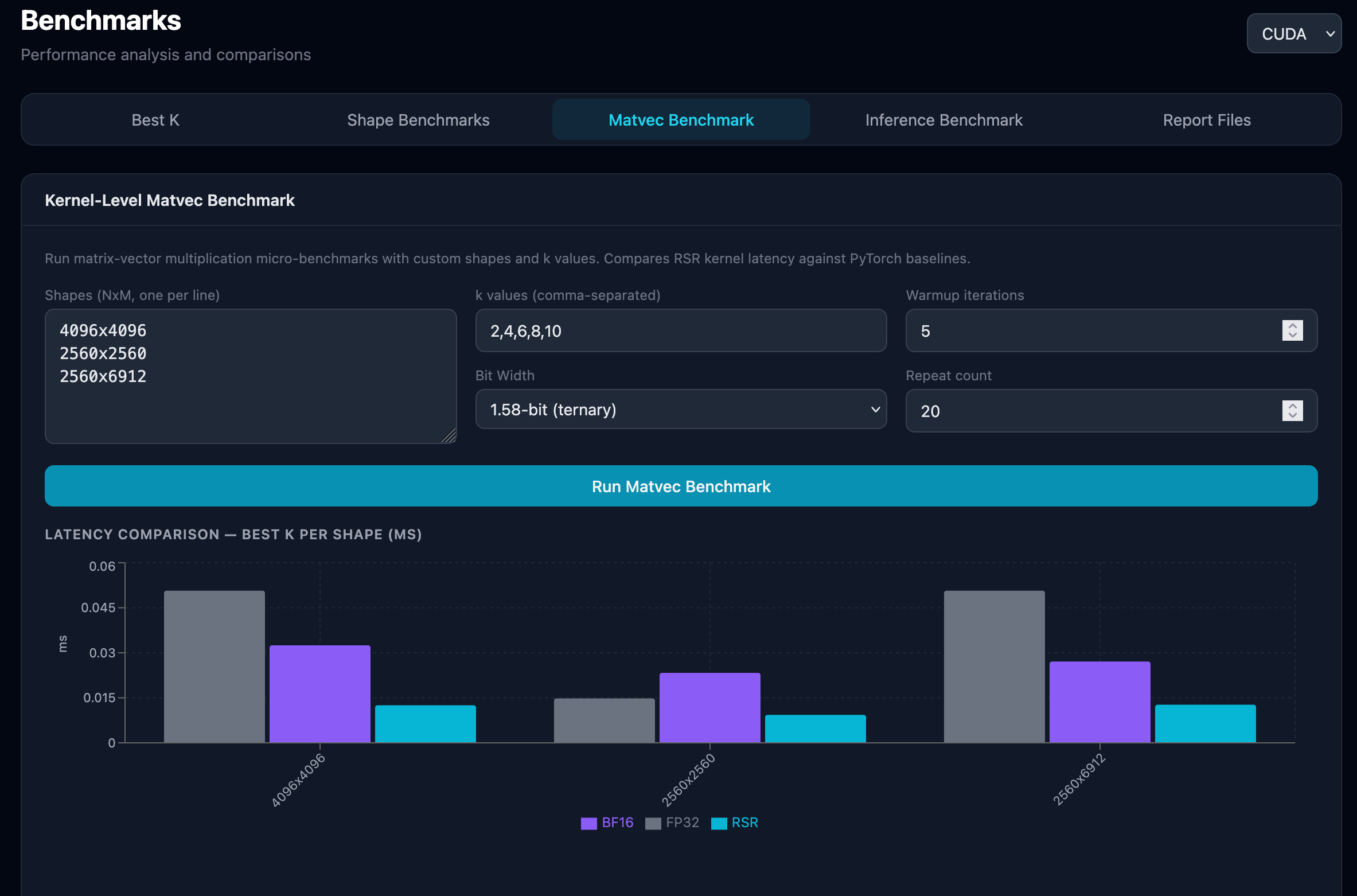}
    \caption{Kernel-level matvec benchmarking.}
    \label{fig:ui_matvec}
  \end{subfigure}
  \hfill
  \begin{subfigure}[t]{0.48\textwidth}
    \includegraphics[width=\textwidth]{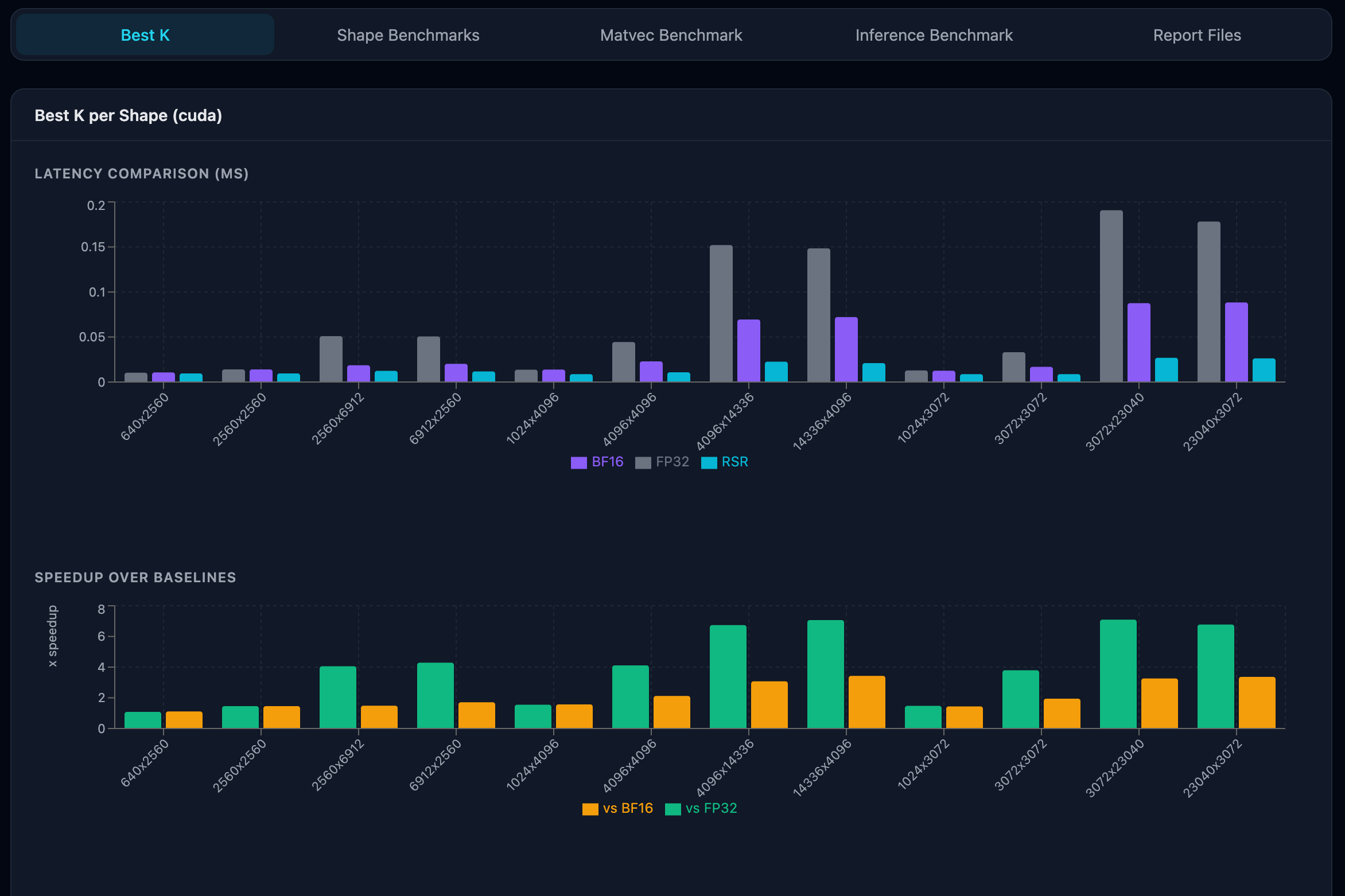}
    \caption{Cross-configuration comparison.}
    \label{fig:ui_bestk}
  \end{subfigure}

  \caption{RSR-core web interface demonstration scenarios.}
  \label{fig:ui_all}
\end{figure*}

\section{Demonstration Scenarios}

During the demonstration, participants interact with RSR-core through an integrated web-based interface that supports model preprocessing, accelerated inference, and kernel-level performance exploration workflows.

\textbf{Scenario 1: Model Onboarding.}
Participants search for a ternary model on the HuggingFace Hub, select it, and configure RSR preprocessing parameters including the block height $k$, multiplier version, and target device. Upon launching preprocessing, the UI displays real-time progress as each layer is processed. Once complete, the model appears in the model list with metadata such as size, layer count, and selected $k$. RSR preprocessing is a one-time cost; the UI makes it accessible to non-experts. Figure~\ref{fig:ui_preprocessing} illustrates the preprocessing interface.

\textbf{Scenario 2: Side-by-Side Inference.}
Participants select a preprocessed model, enter a prompt, and run text generation with both the RSR backend and the HuggingFace \texttt{bfloat16} baseline. The UI displays the generated text alongside performance metrics---tokens per second, generation latency, and model load time---and computes the end-to-end speedup factor. This scenario demonstrates RSR's practical acceleration on real LLM inference. Figure~\ref{fig:ui_inference} shows the inference comparison interface.

\textbf{Scenario 3: Kernel-Level Exploration.}
Participants navigate to the Matvec Benchmark tab, enter custom matrix shapes (e.g., matching a specific model's weight dimensions), select a range of $k$ values and bit-width (1-bit or 1.58-bit), and launch kernel-level micro-benchmarks. The UI produces a bar chart comparing RSR against FP32 and BF16 baselines at the best $k$ per shape, and a line chart showing how RSR latency varies across $k$ values. Participants can explore the parameter sensitivity of RSR on the demonstration hardware, validating that redundancy structure varies across shapes. Figure~\ref{fig:ui_matvec} illustrates the kernel-level benchmarking interface.

\textbf{Scenario 4: Cross-Configuration Comparison.}
Participants switch between CPU and CUDA using the device selector and compare pre-computed shape benchmark results across bit-widths (1-bit vs.\ 1.58-bit). The Best~K tab reveals how the optimal block height shifts across devices and matrix shapes, demonstrating that RSR's advantage is hardware-dependent. The dashboard enables rapid exploration across configurations without re-running experiments. Figure~\ref{fig:ui_bestk} shows the cross-configuration comparison interface.

\bibliographystyle{ACM-Reference-Format}
\bibliography{sample}

\end{document}